# Intelligent Lighting System Using Wireless Sensor Networks

A.A.Nippun Kumaar[1], Kiran.G[2], Sudarshan TSB[3]

Department of Computer Science & Engineering,
Amrita Vishwa Vidyapeetham, School Of Engineering,
Bangalore Campus, India.
[1]nippun05@gmail.com [2]kiran.per.sempre@gmail.com [3]sudarshan.tsb@gmail.com

**ABSTRACT**

*This paper examines the use of Wireless Sensor Networks interfaced with light fittings to allow for daylight substitution techniques to reduce energy usage in existing buildings. This creates a wire free system for existing buildings with minimal disruption and cost.*

## 1. Introduction

Power conservation is no longer just a fashionable expression. It has now become a necessity. Static method of conservation like usage of electrical devices with lower power consumption or scheduled power cuts are not very efficient. This paper proposes a dynamic automated power conservation system which uses wireless sensor networks(WSN). The advantage of using WSN is that this system can be easily installed in already existing buildings where as a wired system will be expensive and difficult to install in the same scenario.

The use of wireless sensor network greatly reduces the size and cost of the system and is suitable for a lighting system.

In the proposed system, there is an array of light sensor nodes which can communicate with a master node(MN), providing information about the light conditions at each sensor node. Based on the feedback information the MN decides which all light sources to control. Once this is decided the MN transmits the data frame to a particular light control node to control the light, which is electrically connected to it.

## 2. Literature Survey

Examined the use of Wireless Sensor Networks interfaced with Dimmable Fluorescent light fittings[1]. Dimmable fluorescent fittings, using modern electronic ballast dimmers are widely fitted to new buildings, to allow for the accurate dimming and control of building lighting[2] F.O'Reilly & J.Buckley. Factoring in natural incident daylight, allows a reduction in the artificial light (daylight substitution), which amounts to savings between 10% and 40%.The DALI light control interface provides a two wire low voltage control bus to allow the addressing and control of individual light fittings[3].

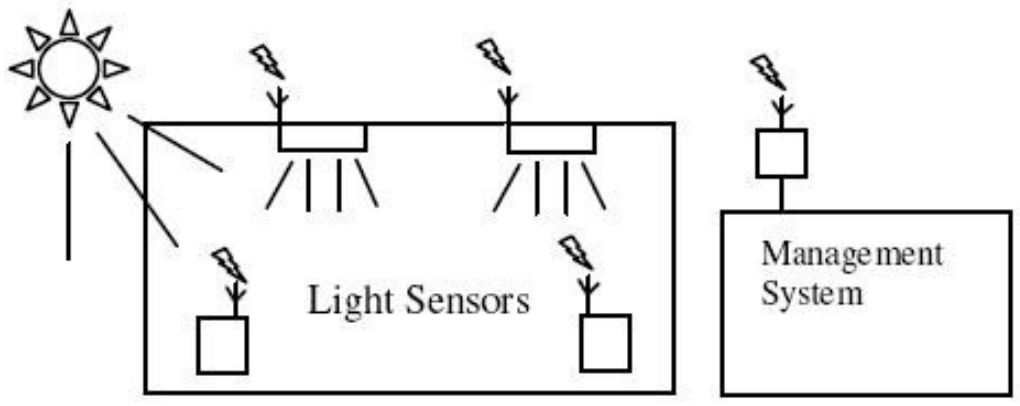

Figure 1: Wireless Daylight Substitution

Figure1 shows a Wireless Sensor Network system which can provide work plane light measurements, and is integrated with a standard building monitoring system, the wireless network controls the dimmable ballast elements, allowing the retrofitting of existing installations without the need to re-cable and with minimal disruption.

The specifications and variations required for work plane lighting, for some sample areas are shown in Table 1, full specifications are available in the CIBSE Lighting Guides[4]. Individual work plane light levels are typically read and forwarded to a facilities management system which can issue control signals to the lighting elements.

| | |
|---|---|
| Filing - Office Work | 300 lux |
| General Office (writing, typing) | 500 lux |
| Fine Painting (Industry) | 750 lux |
| Precision Assembly (Industry) | 1000 lux |

Table 1: Light intensity value for various environment

Even though in some systems human behavior has been considered as a factor and system behavior is based on predictions based on these factors[6][7]. But this paper is directed towards the efficient algorithm design for intelligent lightening system using wireless sensor networks with day light as a important factor.

## 3. Proposed Implementation

In the proposed system, there is no separate base station. One of the nodes will act as the base station. Base station's power is replenishable. Dynamic topology control is done by base station, by periodically ensuring the presence of all nodes and accepting new nodes on the run.

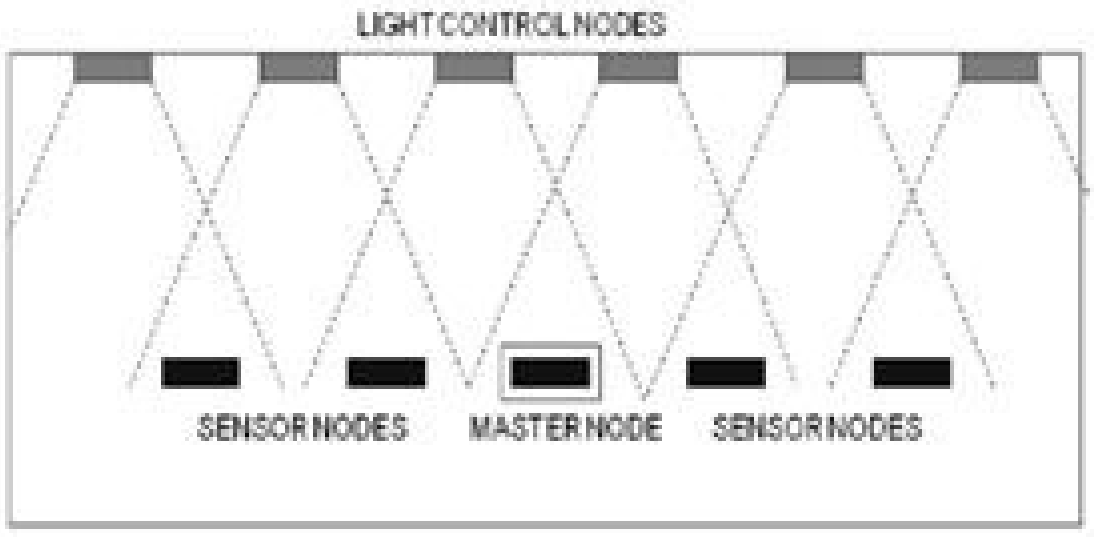

Figure 2: Infrastructure Under Test

As shown in Figure2 there are three kinds of nodes in the network, master node (MN), sensor node (SN), and light control node (LCN). Master node is the one acts as a base station as well as sensor node. Sensor node senses the environment and instructs the light level to the master node. Light control node will respond to the master node by dimming or brightening the light according to the data received. The sensor nodes are placed such that each sensor node ranges to two light ballast. This arrangement will make the light control precise.

# 4. Hardware

Basically the hardware level of this system is classified in two forms, one is in sensor nodes another is in light control node. One of the SN is chosen to be a MN which is loaded with additional control software. Both SN and LCN is controlled by PIC 16F877A controller as shown in Figure4 and Figure5 [5].

## 4.1. Sensor Node

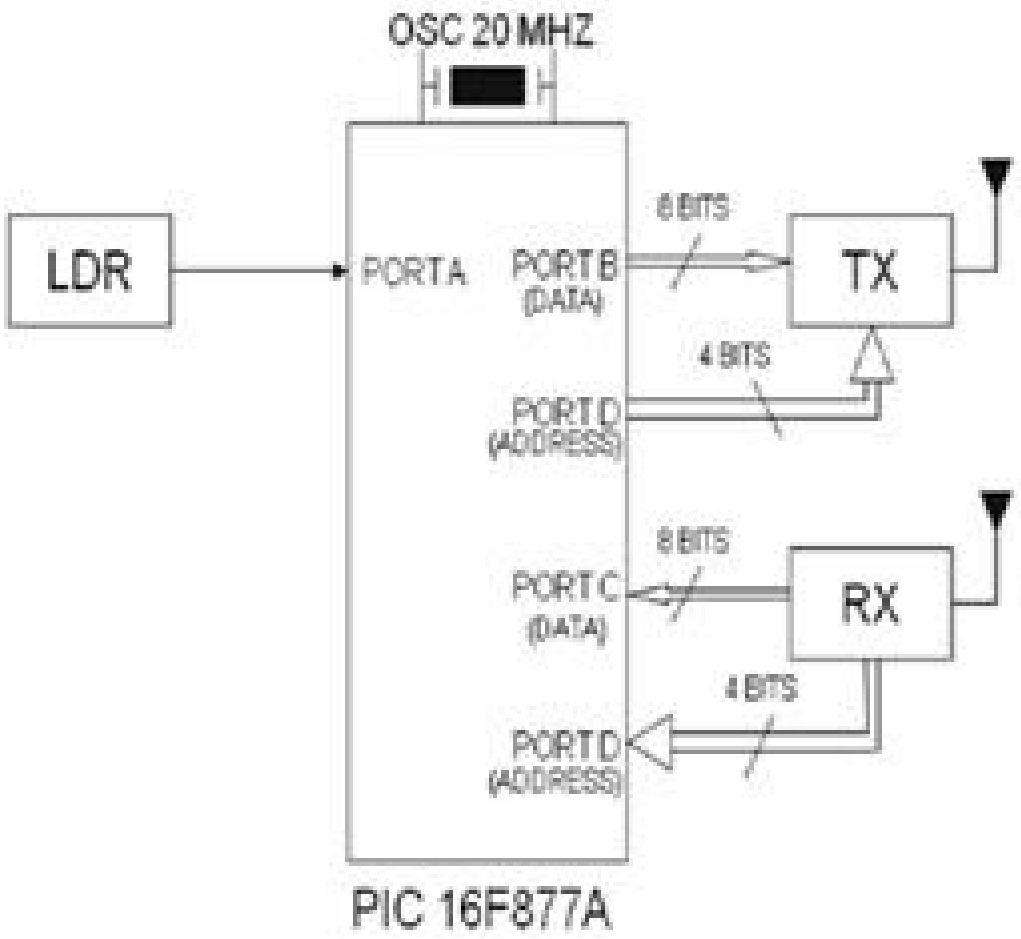

Figure 3: Block Diagram-Sensor Node

Main task of sensor node is to sense the surrounding light level and report to master node. For sensing the light level light dependent resistor (LDR) is interfaced to the controller. As the name suggest resistance of LDR changes when light falls on it. When light increases resistance decreases and vice versa.

The resistance of the Light Dependent Resistor (LDR) varies according to the amount of light that falls on it. The relationship between the resistance RL and light intensity Lux for a typical LDR is

$$RL = \frac{500}{\text{Lux}}\ \text{K}\Omega$$

With the LDR connected to 5V through a R1 K resistor, the output voltage of the LDR is

$$Vo = \frac{5 * \text{RL}}{\text{RL} + \text{R1}}$$

Reworking the equation, we obtain the light intensity

$$\text{Lux} = \frac{2500 * R1}{Vo - 500}$$

LUX -Intensity of light.

Vo -Output voltage from LDR.

R1 -Series resistance connected to LDR

System has a RF transmitter (FS 1000A) and receiver (PCR 2) for wireless transmission and reception. Each node has a pair of Tx and Rx, through this arrangement point to point and broadcast arrangement is possible. Some features of Tx anr Rx is listed below:

- Operating frequency - 315/433 MHz
- Range - 80m
- Data rate - 4KB/s
- Working mode - AM
- Power - 10mW

Transmitter consists of encoder HT 640L. This helps in addressing individual nodes in point to point communication. This allows a maximum of 8bit address and 8bit data frames. This converts parallel transmission of data into serial transmission

## 4.2. Light Control Node

LCN is used to control light intensity according to the received signal. Light controller is nothing but a D/A convertor which will give analog voltage with respect to digital signal. RF Tx and Rx are same as that used in SN.

PIC is used as a controller in both the nodes and plays different role in all the nodes. In sensor nodes A/D convertor of PIC is used to convert LDR voltage into digital voltage, and according to voltage level that has been sensed a data frame is formed and transmitted to MN. In MN the received data is analyzed and data signal is sent to corresponding LCN to control light. MN also maintains three tables MN, LCN and SN table, LCN address table and SN table. In LCN the received signal is analyzed and action is taken accordingly, through light controller.

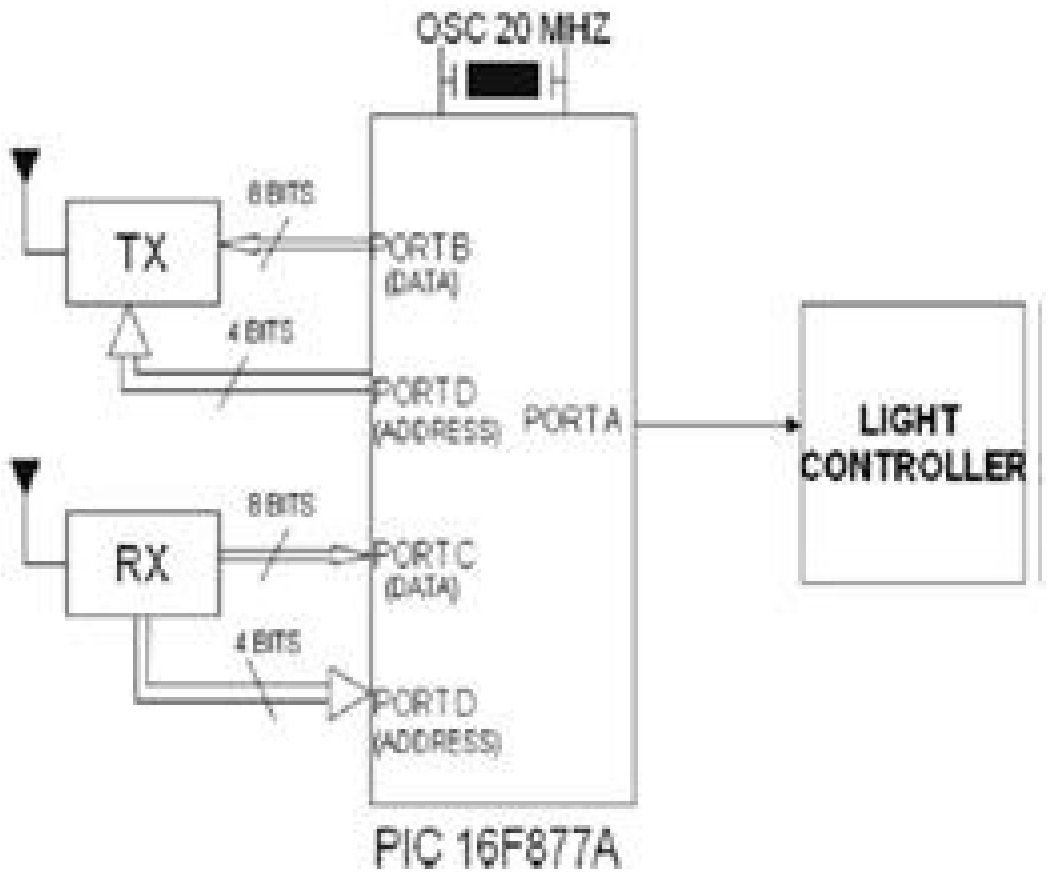

Figure 4: Block Diagram-Light Control Node

In all the three nodes receiver stack is maintained with received data frames. Topology control process is carried out in all the nodes periodically at fixed interval of time.

# 5. Software

The software level of the network is in three forms each in MN, SN and LCN.  There is specially designed frame format for control and data frame transmission.

## 5.1. Frame Format

The frame is designed to be 8bit. Addressing of nodes is carried out both in hardware and software. An address of the node is assigned by the hardware and ID to each node is assigned by software running on MN. The frame is as shown in Figure5, which consists of 2 control bits C1,C2, a topology control bit, a data and acknowledgement bit and 4-bits for assigning address and ID.

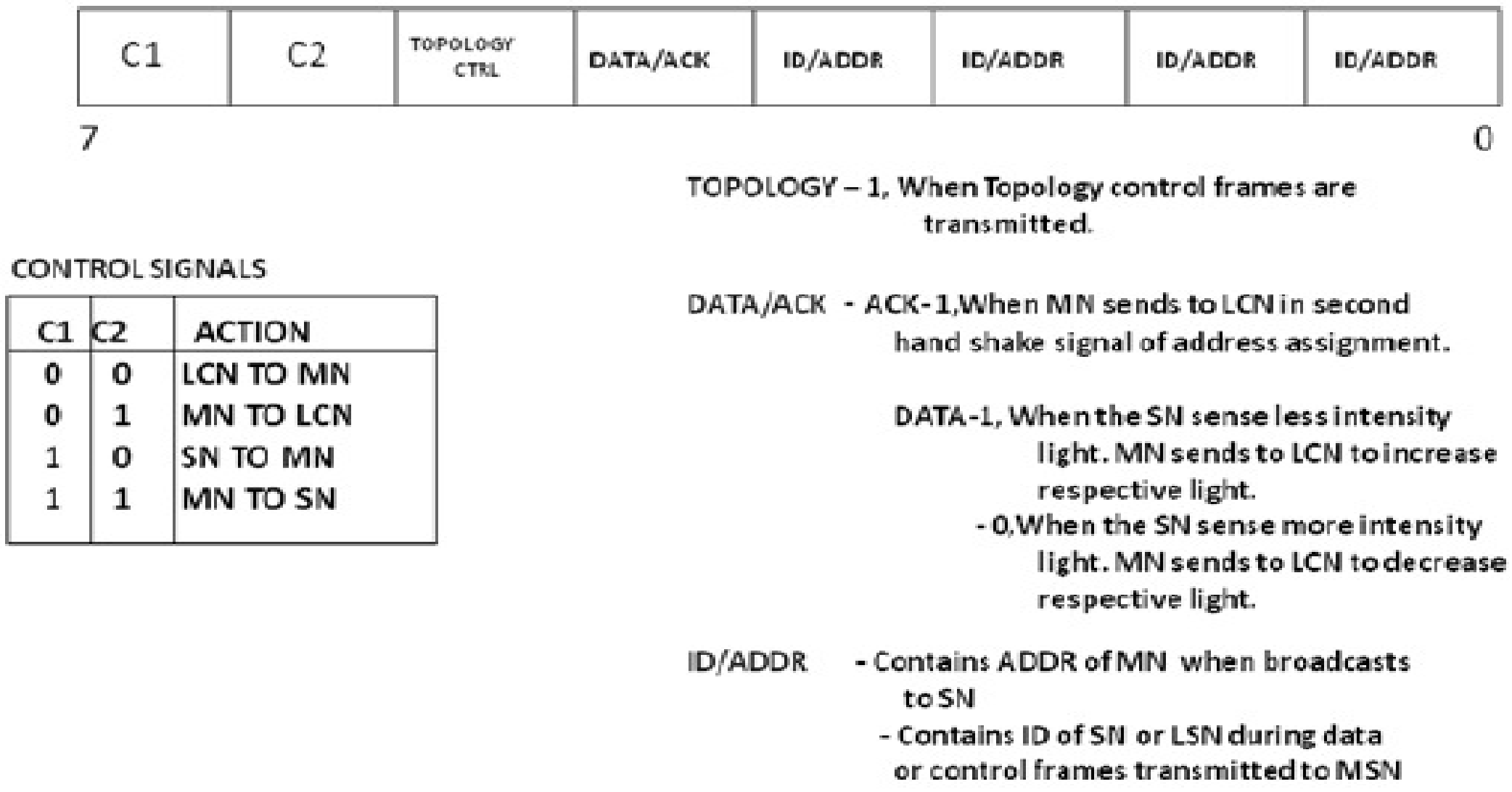

Figure 5: Frame Format

## 5.2. Algorithm

To enable communication between SN and LCN there are three algorithms in the system running in parallel:

- MN Algorithm.
- SN Algorithm.
- LCN Algorithm.

### 5.2.1. MN Algorithm

```
LISTEN any data from LCN
        VERIFY IF the data is
                Invalid frame
                    or
                Address present in LCN table
        END     goto LISTEN
        RECEIVE the data
                Send acknowledgement to LCN
                Assign address to LCN
                Update LCN table
                Send LCN address
        END
        CHECK IF LCN table
                Is full
                  or
                Timer expires
END goto PHASE2
ELSE goto LISTEN

PHASE2: BROADCAST MN address to all SN
        LISTEN2: any acknowledgement from SN
                VERIFY IF the data is
                        Invalid frame
                             or
                       Address present in LCN table
                END     goto LISTEN2
```

```
                RECEIVE the data
                        Update SN table
                END
                CHECK IF SN table
                        Is full
                          Or
                        Timer expires
                END goto PHASE3
        ELSE goto LISTEN2
END

PHASE3: SEND increment data to LCN
                LISTEN request
                        For decrease light from SN
                        Check for five request from same SN
                END
                UPDATE SN table with LCN
                             address
                CHANGE LCN address
                GOTO SEND
          END repeat for all LCN
         CHECK IF for
                For all LCN mapped
                        Or
                Timer expires
         END goto NORMAL
ELSE goto PHASE3
NORMAL: LISTEN any data from SN
        VERIFY the data for
                Invalid frame
        END goto LISTEN
        READ data
                For SN ID
                Check ti INC or DEC light
```

Get the LCN address from SN table

**END**

**SEND** the control signal to that LCN

**END** goto NORMAL

This has 4 phases. In the first phase address are assigned for LCN's. Whenever a frame from LCN is released it is updated in the LCN table. In the second phase MN broadcasts its own address and wait for the SN to reply. Replies from SN are used to update the SN table. In the third phase a mapping is done between LCN and SN i.e. a table is updated that maps the SN, controlled by a particular LCN. This is done by selectively brightening the lighting source controlled by a SN to the maximum value of the LCN. In the fourth phase, which signifies a normal operation SN frames are received by MN and "increase or decrease light" frames are sent to the LCN for finer control of luminance.

### 5.2.2. SN Algorithm

**LISTEN** any broadcast data is received

**VERIFY IF** the data

Is invalid

**END** goto LISTEN

**READ** the frame

Take the MN address

Set it as its TX address

Send ACK as its ID

**END**

**LDR SENSE**: check the light intensity level

**VERIFY IF** the value

Is higher or lesser than threshold

Send the data to MN accordingly

**END** goto LDR SENSE

**END** goto LDR SENSE

Here, SN waits for the MN broadcast. Once it receives the address, it configures its transmitter to a permanent address. As acknowledgement it sends its own ID. During normal operation it constantly senses the light and whenever the light goes below or above the threshold, it will send "increase os decrease light" frame.

#### 5.2.3 LCN Algorithm

```
SEND its ID to MN
LISTEN to any data from MN
    VERIFY IF the data
        Is invalid
    END goto LISTEN
    READ the frame
        Extract ID
        Check with its ID
        Goto LISTEN2 if ID is same
        Or discard data
    END
END

LISTEN2 to any data from MN
    VERIFY IF the data
        Is invalid
    END goto LISTEN2
    READ the frame
        Extract the address
        Set the address as its RX address
    END
END

NORMAL: any data from MN
    VERIFY IF the data
        Is invalid
    END goto NORMAL
    READ the received data
        Check for INR or DCR light
        Control the light accordingly
    END goto NORMAL
END
```

Initially LCN will send its own ID to MN. MN will reply receiver's address allotted to it through three way handshaking. LCN will configure its receiver with this address. During normal operation, it listens for MN frame. When it receives "increase or decrease light" MN frames, it controls the luminance accordingly.

# 6. Results

Experimental Setup:

1. A room with five lights, four at corners and one at the middle.
2. Consider the intensity required in the room should be 400Lux and the light should be lit up for 12 Hrs/day (6Hrs day & 6Hrs night).
3. The tube light used will consume 40W of power.
4. In normal system all light should glow in full intensity therefore consumes 40W each.
5. In the proposed system all lights in corner needs only 50% of the power in day time.

Table 2 shows the comparison for the given setup between normal system and the proposed system. The savings in energy consumed for the given setup is observed to be 14400 Wh/Month.

| Normal | | Power | Hrs Used | No. Of. Light | Energy Consumed Per day | Total Energy Consumption |
|---|---|---|---|---|---|---|
| System | Day | 40W | 6 | 5 | 1200Wh | 2400 Wh/ Day<br>72000 Wh/ Month |
| | Night | 40W | 6 | 5 | 1200Wh | |
| Proposed system | Day | 20W | 6 | 4 | 480Wh | 1920 Wh/ Day<br>57600 Wh/ Month |
| | | 40W | 6 | 1 | 240Wh | |
| | Night | 40W | 6 | 5 | 1200Wh | |

Table 2: Result analysis

# 7. FUTURE IMPLEMENTATION

So by adding PIR sensor which will detect human presence alone will add more intelligence to the system and further helps in reduction of power by selectively dimming or switching off some light sources and thus keeping average power consumption constant.

# 8. CONCLUSION

Through this system we introduce one more way of "Going Green". Installing wired devices for the same purpose may not be cost efficient and can even be counter productive. Our device is easy to install and manage and thus more appealing.

Compared to the original paper our system is more scalable and flexible. Runtime addition of nodes is possible and better power efficiency can be obtained. Usage of custom control equipment reduces the cost as well. Thus adding to the appeal.

**Authors**

**A.A. Nippun Kumaar** is currently a graduate student pursuing Masters in Embedded Systems from Amrita Vishwa Vidyapeetham, School Of Engineering, Bangalore Campus. He completed his Diploma in Electronics and Communication Engineering with First Class Honors in Thiyagarajar Polytechnic College, Salem, Tamilnadu, India in 2003. He completed his Bachelors degree, B.E. in Electronics and Communication Engineering with First Class Honors in Sona College of Technology, Salem, Tamilnadu, India in 2006. His areas of interest are Wireless Sensor Networks and Robotics.

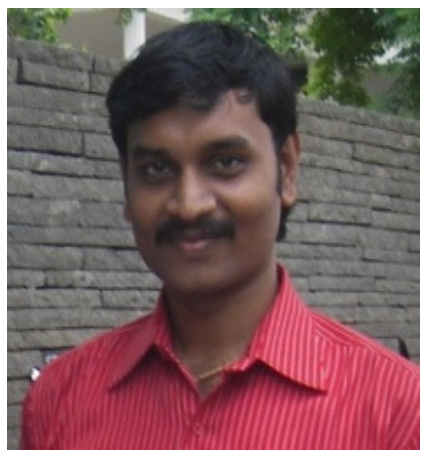

**Kiran. G** is currently a graduate student pursuing Masters in Embedded Systems from Amrita Vishwa Vidyapeetham, School Of Engineering, Bangalore Campus. He completed his graduation in Electronics and communication from Institute of science and Technology, ErnaKulam, Kerala, India in 2007, followed by a PG Diploma in Embedded System Design. He started developing interest in embedded systems during his graduation and his other areas of interest are unix system programming, digital signal processing.

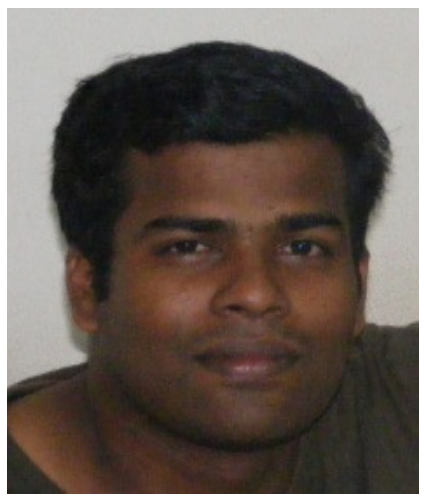

**Dr. Sudarshan TSB** is a Professor, Amrita Viwa Vidyapeetham, School of Engineering, Bangalore Campus., India. Earlier he served as Assistant Professor and then Head of the department of Computer Science & Engineering, BITS, Pilani for 13 years. He completed his Doctrate from BITS, Pilani, India in the area of computer networks. His research Interests are Multicore Computing and Wireless Networks.

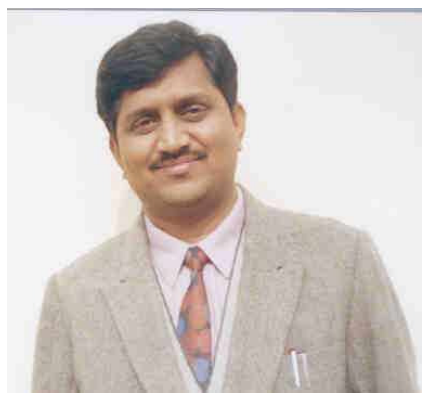